\documentclass[conference]{IEEEtran}
\IEEEoverridecommandlockouts
\usepackage{cite}
\usepackage{amsmath, amssymb, amsfonts, subcaption, bm, amsthm}
\usepackage{graphicx}
\usepackage{textcomp}
\usepackage{xcolor}

\newcommand{\rev}[1]{\textcolor{black}{#1}}
\newcommand{\revi}[1]{\textcolor{black}{#1}}

\newtheorem{lemma}{Lemma}
\newtheorem{theorem}{Theorem}
\newtheorem{corollary}{Corollary}
\newtheorem{remark}{Remark}

\def\BibTeX{{\rm B\kern-.05em{\sc i\kern-.025em b}\kern-.08em
    T\kern-.1667em\lower.7ex\hbox{E}\kern-.125emX}}
\begin{document}

\title{Expanding Over-the-Air Computation\\with Frequency Modulations
\thanks{This work is supported by the Spanish National projects AEI/10.13039/501100011033 SOFIA-SKY (PID2023-1473050B-C32).}
}

\author{\IEEEauthorblockN{Marc Martinez-Gost\IEEEauthorrefmark{1}\IEEEauthorrefmark{2}, Ana Pérez-Neira\IEEEauthorrefmark{1}\IEEEauthorrefmark{2}\IEEEauthorrefmark{3}, Miguel Ángel Lagunas\IEEEauthorrefmark{2}}
\IEEEauthorblockA{
\IEEEauthorrefmark{1}Centre Tecnològic de Telecomunicacions de Catalunya, Spain\\
\IEEEauthorrefmark{2}Dept. of Signal Theory and Communications, Universitat Politècnica de Catalunya, Spain\\
\IEEEauthorrefmark{3}ICREA Acadèmia, Spain\\
\{mmartinez, aperez, malagunas\}@cttc.es}
}

\maketitle

\begin{abstract}
In this study we introduce Logarithmic Frequency Shift Keying (Log-FSK), a novel frequency modulation for over-the-air computation (AirComp). Log-FSK leverages non-linear signal processing to produce AirComp in the frequency domain, this is, the maximum frequency of the received signal corresponds to the sum of the individual transmitted frequencies.
The demodulation procedure relies on the inverse Discrete Cosine Transform (DCT) and the extraction of the maximum frequency component.
Log-FSK enables the computation of functions beyond the sum by incorporating nomographic function representation.
Furthermore, unlike existing AirComp modulations, Log-FSK allows to compute several functions in a single transmission.
We evaluate the capabilities of the scheme in an additive white Gaussian noise (AWGN) and flat-fading channels.
To demonstrate its practicality, we present specific applications and experimental results showcasing the effectiveness of Log-FSK AirComp within linear Wireless Sensor Networks (WSN). Our numerical results show that Log-FSK outperform linear analog modulations in terms of MSE and power consumption.
\end{abstract}


\section{Introduction}
The goal of task-oriented communications is to design protocols that assist the completion of a task \cite{Gunduz22}. Unlike traditional communication systems, the objective is no longer to convey information reliably but to accomplish the specific target. Over-the-air computing (AirComp) has gained attention in the context of distributed function computation \cite{Nazer2007}.
The key behind AirComp is designing codes considering the communication and computation simultaneously. Specifically, AirComp exploits that node identity is irrelevant for many computations (e.g., statistics) and that the additive nature of the communication channel can be used for simultaneous transmission in time and frequency, without the need of orthogonal resource allocation.
In this respect, AirComp is suitable for next generation multiple access (NGMA) schemes, as it removes the orthogonality of resource blocks and ensure that these can be utilized in the most effective manner. Moreover, AirComp provides high spectral efficiency because users transmit using the same bandwidth, as well as low latency, since all transmissions are concurrent.

Nonetheless, most of the literature in AirComp assumes a baseband signal model \cite{sahin22}, which ignores the need of a modulation that matches the additive nature of the communication channel. As analyzed in \cite{perez2024waveforms}, these works implicitly assume a linear analog modulation (i.e., either single or double sideband) where the amplitude carries the information. On the other hand, the consideration of angular modulations is of wide interest due to their implementation and performance advantages. However, their applicability for AirComp is not straightforward due to their nonlinear nature, i.e., $e^{ja}+e^{jb}\neq e^{j(a+b)}$. In general, using any modulation for AirComp is not possible, as it requires to match the linear additive property of the channel.
This topic has been overlooked in the literature 
and most of the progress has been devoted to the use of existing modulations for specific computations in AirComp, that are not general enough to encompass more general processing schemes. 

In this work we propose Logarithmic Frequency Shift Keying (Log-FSK), a novel waveform for AirComp that relies on digital FSK. To overcome the concerns of linear analog modulations, Log-FSK carries information only in the frequency, not in the amplitude of the signal. The modulation is designed so that a nonlinear processing at the receiver yields the sum of the modulated information in the frequency. In other words, the resulting frequency is the sum of individual transmitted frequencies. While the Kolmogorov superposition theorem \cite{Kolmogorov} and the nomographic representation of functions \cite{GOldenbaum13} have been widely used in the AirComp community to compute functions in a distributed fashion, Log-FSK is the first modulation that has been designed using a nomographic representation.
Besides, we highlight the fact that FSK modulations are currently used in communication networks (e.g., LoRa modulation \cite{Chiani2019}) and that Log-FSK does not require changing the overall architecture, but append additional blocks to standard communication systems.
\rev{
Log-FSK addresses a specific yet crucial scenario in AirComp: low-density networks that require high-resolution function computation, which existing AirComp techniques do not adequately support. While classical digital schemes could be employed given the limited number of transmitters, Log-FSK offers distinct advantages tailored for AirComp. First, it enhances computational efficiency by seamlessly integrating computation within the communication process, reducing the processing burden at both the transmitter and receiver. Second, it significantly lowers latency, as its modulation inherently enables direct aggregation of transmitted values, eliminating additional decoding or computation steps. These properties make Log-FSK particularly valuable in real-time and low-latency applications where efficient computation and communication are equally critical.
}

We first present the waveform in time and frequency domains, and show how, despite not being sinusoidal, it can be implemented with linear oscillators. 
Then, we theoretically prove the additive nature of Log-FSK in the frequency domain and propose the demodulation procedure to recover the AirComp computation.
We evaluate the performance of Log-FSK in an additive white Gaussian noise (AWGN) for different number of simultaneous transmitters and extend the results for flat-fading channels.
As Log-FSK is a digital modulation, we provide the theoretical performance of Log-FSK in terms of error probability and the equivalent mean squared error (MSE).
Regarding the computation of functions, we limit the study to a two-user scenario and exploit the nomographic function representation to show how Log-FSK can be used to compute generic functions beyond the sum. Furthermore, we show how Log-FSK allows to compute several functions in a single transmission, which cannot be done by existing modulations.

This journal paper presents an extension of our previous work \cite{gost24}, where we initially introduced Log-FSK. The main contributions of this paper are described in the following:
\begin{enumerate}
    \item We propose Log-FSK, a digital frequency modulation for AirComp. We characterize the waveform in both time and frequency domains, as well as an implementation that relies on linear oscillators.

    \item We provide the theoretical performance of Log-FSK in terms of error probability and MSE in an AWGN and fading channels. We also compare it with linear AirComp, implemented with double sideband (DSB) modulation, in terms of MSE for different SNR levels. 
    
    \item We provide a power control strategy to minimize the MSE of AirComp. We thoroughly analyze the power consumption in AirComp for both linear modulations (DSB) and frequency modulations (Log-FSK), for different data distributions and functions.

    \item We provide a variety of use cases and setups under which the system can be implemented for non-massive access.

    \item We adopt the nomographic representation to extend the function computation beyond the sum.
\end{enumerate}

The remaining part of the paper proceeds as follows:
Section II starts with a review of the literature of modulations for AirComp.
Section III introduces the Log-FSK waveform, characterizes its spectrum, proposes an implementation with linear oscillators and present the modulation and demodulation procedures.
Then, section IV analyses the performance of Log-FSK in an AWGN channel, and extends its applicability over fading channels. Section V proposes specific computation of functions that can be computed with Log-FSK beyond the sum, and Section VI complements the theoretical analysis with experimental results. Section VII concludes the paper.

\textit{Notation}: $\mathbb{N}$, $\mathbb{R}$ and $\mathbb{C}$ stand for the set of natural, real and complex numbers, respectively; lowercase and uppercase bold symbols correspond to vectors and matrices, respectively;
$\mathbf{s}[i]$ corresponds to the $i$th entry of vector $\mathbf{s}$, $diag(\mathbf{s})$ is a diagonal matrix with the entries of $\mathbf{s}$ and $\delta[n]$ is the Kronecker delta.

\section{Literature Review}
The inception of computing functions over multiple access channels is traced back to \cite{Nazer2007}, in which the authors show that the source-channel separation theorem does not hold, even when the data is independent among sources. For general networks (i.e., not point-to-point) using digital codes (i.e., transmitting bits) is generally suboptimal, while simple analog architectures (i.e., analog uncoded data) using joint source-channel communication perform optimally \cite{Gas06}.
Despite the advances in information theory, these results do not provide a procedure (e.g., coding scheme) to compute generic functions, such as the mean or the maximum.
In this respect, the authors in \cite{Jeon14} propose a computation code for computing weighted sums and the type function (i.e., the histogram).
Later, \cite{GOldenbaum13} established the connection between AirComp and nomographic functions. These functions can be computed in a distributed fashion with an appropriate preprocessing at each transmitter, simultaneous transmission over the channel and a postprocessing at the receiver.

From a physical layer perspective, the choice of modulations for AirComp remains an open research topic \cite{perez2024waveforms}. The most natural way of implementing AirComp is using amplitude modulations and direct aggregation (DA), this is, all users transmitting using the same time and frequency slot. This is the case of broadband analog aggregation (BAA), where the analog information is carried in the amplitude of orthogonal frequency division multiplexing (OFDM) subcarriers \cite{baa}. However, linear modulations are susceptible to noise and channel impairments, and are not currently deployed in communication systems. In a different vain, a minimum shift keying (MSK) is proposed in \cite{Katti07} for analog network coding. However, this specific domain constrains the design to two users only.
In \cite{Sigg12} the information is encoded in the mean of a Poisson distribution, which is used to generate and transmit bursts. When several transmitters generate bursts, the process observed by the receiver results in a Poisson distribution whose mean is the sum of the individual means.

While some authors propose the use of frequency modulations, these resort on the so-called type-based multiple access (TBMA), which still carries information in the amplitude of the carrier.
In TBMA, orthogonal resources are assigned to different measurements, so that AirComp only happens when several users modulate the same information and transmit using the same resource \cite{mergen06}.
In \cite{sahin21} a ternary FSK is proposed to encode the sign of the gradient in a federated learning system (i.e., positive, zero or negative); in \cite{martinez23} and \cite{Gost_Globecom23} TBMA is used for distributed function computation and federated learning, respectively. This time, the data is assigned to orthogonal carriers in $M$-ary FSK. While all these schemes employ frequency modulations, TBMA still relies on the amplitude of the demodulated data to estimate the function value. In the ternary scheme, a majority voting scheme is used to select the final sign of the gradient, while in the $M$-ary case, the amplitudes of all frequency bands are used to compute the average of the transmitted data. In this respect, the performance of the scheme is not fully characterized by the error probability, as it happens in frequency modulations. 
To overcome these issues, the authors in \cite{razavikia23} recently proposed designing digital constellations for AirComp. At the receiver side, the aggregation of transmitted symbols corresponds to a new symbol associated to a unique function value.
However, constellations are not dense and change for every function and number of users.
We refer the reader to \cite{perez2024waveforms} to an overview of the existing analog and digital modulations and techniques for AirComp.





Inspired by the optimality of uncoded transmission and the ubiquity of digital modulations, we propose a digital modulation for analog data. In this respect, the analog data is quantized and the only loss is with respect to the quantization noise, that can be reduced using more quantization levels. Moreover, driven by the benefits of frequency over amplitude modulations, the proposed modulation is frequency-based.

\section{Waveform Design}
\label{sec:waveform_design}

\subsection{Log-FSK modulation}

We first construct an $M$-ary frequency shift keying ($M$-FSK), where each user $k=1,\dots,K$ maps its measurement $m_k\in[0,N-1]\subset{\mathbb{N}}$ to a specific frequency. Then, we introduce a logarithmic transformation over the waveform, resulting in the following discrete-time signal:
\rev{
\begin{align}
    x_k[n] &= \log\left(\beta\sqrt{\frac{2}{N}}\cos\left(\frac{\pi(2m_k+1)}{2N}n\right)+1\right)\nonumber\\
    &=\log\left(\beta\cos_{m_k}[n]+1\right),
    \label{eq:logos_modulation}
\end{align}
}
for $n=0,\dots,N-1$, where $N=M$ is the number of samples and $\log(\cdot)$ is assumed to be the natural logarithm without loss of generality. \revi{Parameter $0<\beta<\sqrt{N/2}$ is a scaling factor that prevents the argument of the logarithm from reaching negative values and controls the signal power.} The mean of $x_k[n]$ may be subtracted so that the signal has no DC component.
In \eqref{eq:logos_modulation} we defined
\begin{equation}
    \cos_{m_k}[n]=
    \sqrt{\frac{2}{N}}\cos\left(\frac{\pi(2m_k+1)}{2N}n\right)
    \label{eq:cosine}
\end{equation}

\begin{figure}[t]
\centering
\includegraphics[width=\columnwidth]{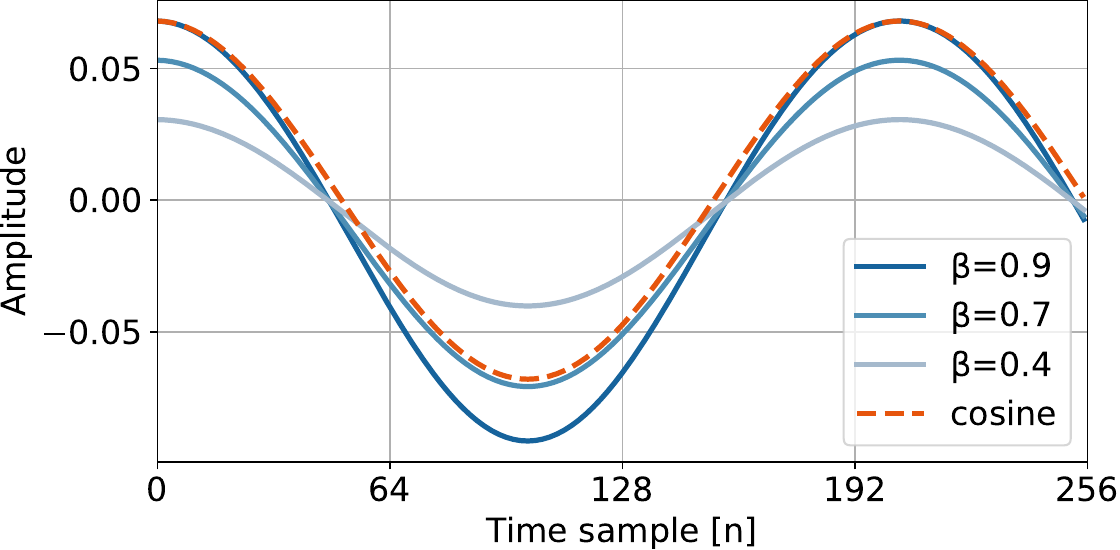}
\caption{\rev{Log-FSK waveform for different $\beta$ and a cosine waveform at frequency $m=5$.}}
\label{fig:logos_waveform}
\vspace{-0 pt}
\end{figure}

These cosines correspond to the DCT basis (see \cite{Gost23}), and we refer to the modulation in \eqref{eq:logos_modulation} as the Logarithmic FSK (Log-FSK). This design will be justified in Section \ref{sec:modulation}. Figure \ref{fig:logos_waveform} shows the waveform along with a corresponding cosine at the same frequency ($m=5$, $N=256$). The proposed signal retains the same periodicity as the cosine due to the monotonicity of the logarithm. However, unlike the cosine, the range is not symmetric.

\subsection{Spectrum of Log-FSK}
To interpret \eqref{eq:logos_modulation} in the frequency domain, we identify its resemblance with the cepstrum \cite{Opp89}. The cepstrum is the inverse transform of the logarithm of the spectrum. Conversely, the Log-FSK waveform corresponds to the logarithm of a time signal. Thus, we will apply duality of the Fourier transform of the cepstrum.
The spectrum of the Log-FSK waveform is presented in Theorem \ref{thm:spectrum}.

\rev{
\begin{lemma}
\label{thm:mean}
The signal $x_k[n]$ in \eqref{eq:logos_modulation} has a mean value of
\revi{
\begin{equation}
    \bar{x}_k=
    \frac{1}{N}\sum_{n=0}^{N-1}x_k[n]=
    \log\left(
    \frac{1+\sqrt{1-2\beta^2/N}}{2}
    \right)
    \label{eq:mean_signal}
\end{equation}
}
\begin{proof}
The proof is presented in the Appendix \ref{app:proof_lemma_mean}.
\end{proof}
\end{lemma}
}

\begin{theorem}[Spectrum of Log-FSK]
\label{thm:spectrum}
The signal $x_k[n]$ in the frequency domain is
\rev{
\begin{equation}
    X_k[\ell]=\bar{x}_k\delta[\ell]-\sum_{i=1}^{\infty}\frac{(-1)^{i}}{i}\left(\frac{1}{2N}\right)^i
    \delta[\ell-im_k],
    \label{eq:spectrum}
\end{equation}
}
for $\ell>0$, where $\ell$ is the discrete frequency variable.
\begin{proof}
The proof is presented in the Appendix \ref{app:proof_thm_spectrum}.
\end{proof}
\end{theorem}

\begin{corollary}[]
Theoretically, Log-FSK has an infinite spectrum. 
Nonetheless, the amplitudes decay at an exponential rate as $\left(2N\right)^{-i}/i$.
\end{corollary}

Figure \ref{fig:logos_spectrum} shows the Log-FSK waveform (left) generated with $m=5$ and $N=256$. The corresponding spectrum (right) shows peaks at multiples of $m$ and the amplitudes present an exponential decay. There is no DC component because the mean has been previously subtracted.

\subsection{Waveform implementation}
\label{sec:waveform_implementation}
Implementing this waveform in hardware may seem cumbersome due to the nonlinearity introduced by the logarithm. Nonetheless, we take advantage of the Discrete Cosine Transform (DCT) for function approximation to implement it using linear oscillators. The DCT of \eqref{eq:logos_modulation} is
\begin{equation}
F_k[\ell] = g_\ell\sum_{n=0}^{N-1} x_k[n] \cos\left(\frac{\pi \ell(2n+1)}{2N}\right),
\label{eq:dct}
\end{equation}
for $\ell=0,\dots, N-1$, where $g_0=1/\sqrt N$ and $g_\ell=\sqrt{2/N}$ otherwise. The $F[\ell]\in\mathbb{R}$ are termed the DCT coefficients and correspond to $X_k[\ell]$. We define $\tilde{x}_k[n]$ as the inverse DCT of \eqref{eq:dct} using only $L$ coefficients:
\begin{align}
\tilde{x}_k[n] =& \sum_{\ell=0}^{L-1} g_\ell F_k[\ell] \cos\left(\frac{\pi \ell(2n+1)}{2N}\right)\nonumber\\
=& \sum_{\ell=0}^{L-1} g_\ell X_k[\ell] \cos\left(\frac{\pi \ell}{N}n + \phi_\ell\right),
\label{eq:idct}
\end{align}
where we define the phase $\phi_\ell=\pi \ell/2N$. Notice that $\tilde{x}_k[n]=x_k[n]$ for $L=N$ and $\tilde{x}_k[n]\approx x_k[n]$ for $L<N$. From the perspective of signal processing, the DCT approximation is appropriate because the largest $L$ coefficients retain most of the energy (see \cite{Gost23} for all the benefits of using the DCT in this context). Furthermore, from communication view, \eqref{eq:idct} allows to implement the Log-FSK waveform using only linear oscillators and bypass the logarithmic implementation in the radio frequency domain.

\begin{figure}[t]
\centering
\includegraphics[width=\columnwidth]{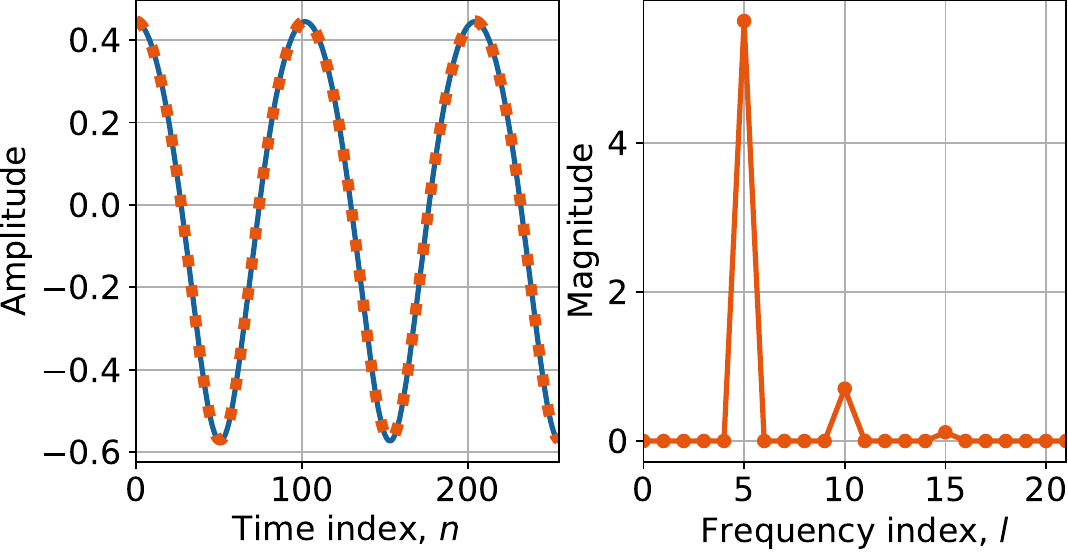}
\caption{Spectrum of Log-FSK: (Left) Proposed waveform ($m_k=5$) and approximation with $L=3$ coefficients (dashed); (right) Spectrum of Log-FSK or magnitude of the DCT coefficients.}
\label{fig:logos_spectrum}
\vspace{-0 pt}
\end{figure}

Figure \ref{fig:logos_spectrum} (right) shows the spectrum of Log-FSK, which correspond to the DCT coefficients. The left plot also shows the corresponding reconstruction using 3 coefficients ($\ell=5, 10 \text{ and } 15$), for which up to $99.998\%$ of the energy is retained. The DCT representation allows to bound the bandwidth of the transmitted signal, which can be approximate as $3m/4T$, where $T$ is symbol period \cite{Gost23}.

\begin{remark}
The number of DCT coefficients determines a trade-off between the MSE in signal approximation and transmission bandwidth. At some point, increasing $L$ only expands the communication bandwidth, at the expenses of a negligible gain in signal approximation.
\end{remark}

\subsection{Over-the-air computation}
\label{sec:modulation}
Consider $K=2$ users modulating their respective measurements, $m_1$ and $m_2$, as in \eqref{eq:logos_modulation}. For simplicity, consider an ideal noiseless multiple access channel (MAC) and both users transmitting concurrently and perfectly synchronized in time. Upon reception, the receiver computes the exponential over the received signal as
\begin{align}
    r[n] &=\exp{\left(x_1[n]+x_2[n]\right)}\nonumber\\
    &=
    \frac{\beta^2}{2}\sqrt{\frac{2}{N}}\left(\cos_{m_1+m_2}[n] + \cos_{|m_1-m_2|}[n]\right)\nonumber\\
    &\,\quad
    +\beta\cos_{m_1}[n]+
    \beta\cos_{m_2}[n] +1,
    \label{eq:FM_AirComp}
\end{align}
\rev{where the trigonometric identity of the product of cosines is expressed as cosine of the sum and cosine of the difference.} The first term in \eqref{eq:FM_AirComp} reveals that AirComp using frequency modulations is, at least, theoretically possible. \rev{In Sec. \ref{sec:Communication} we will generalize this result for flat-fading channels, and an arbitrary number of transmitters.}

The key move for achieving AirComp using frequency modulations is to transform the additive channel into a multiplicative channel, which ultimately results in the addition of frequencies. This is because the product of cosines is equivalent to a cosine of the sum and a cosine of the difference. Log-FSK is conceived applying the nomographic representation,
\begin{equation}
    f(x_1,\dots,x_K)=
    \psi\left(\sum_{k=1}^K \varphi(x_k)\right),
    \label{eq: nomographic}
\end{equation}
where $\varphi$ are the preprocessing or inner functions, and $\psi$ is the postprocessing or outer function. In a communication setting, the preprocessing is computed at each transmitter in a distributed fashion, the sum is performed by the MAC and the postprocessing is computed at the receiver. The multiplication of signals is obtained with $\varphi(\cdot)=\log(\cdot)$ and $\psi(\cdot)=\exp(\cdot)$, where these transformations are applied to the waveforms. In this respect, Log-FSK is the first design that exploits the superposition theorem to design waveforms for AirComp.


\subsection{Log-FSK demodulation}
When the frequency of the sum in \eqref{eq:FM_AirComp} is guaranteed to belong to the original domain, i.e., $\Sigma=m_1+m_2\in[0,N-1]$, it is easy to see that the $\Sigma$ corresponds to the maximum frequency of \eqref{eq:FM_AirComp}. The extension to $K$ users under the same constraint is straightforward.

\begin{figure}[t]
\centering
\includegraphics[width=\columnwidth]{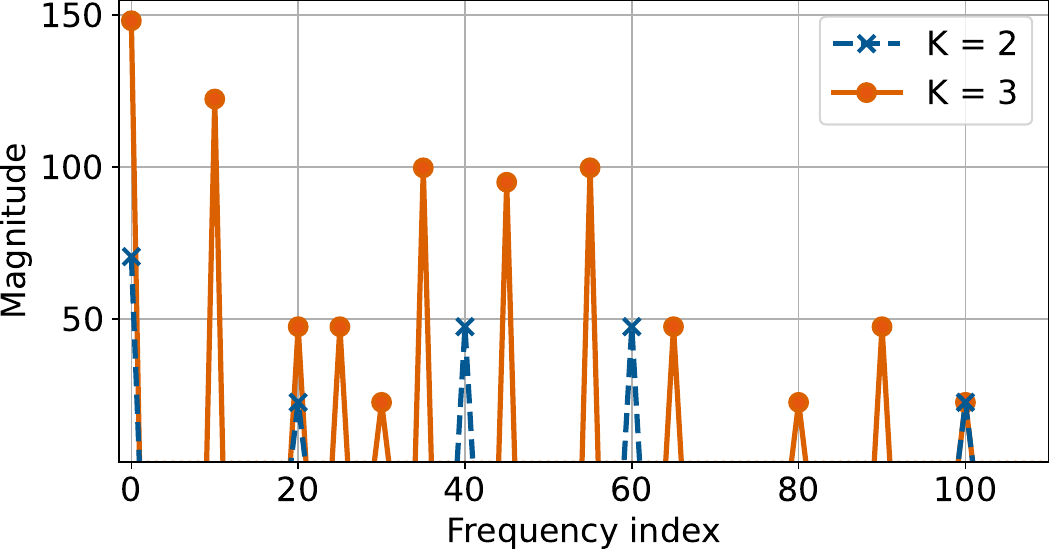}
\caption{Signal after demodulation for $K=\{2,3\}$. The modulated information is $[40, 60]$ and $[10, 35, 55]$, respectively. In both cases the maximum frequency is located at the sum of information, this is, $100$.}
\label{fig:logos_demodulation}
\vspace{-0 pt}
\end{figure}

The previous result allows to design the demodulation process, since it requires detecting the maximum frequency. Provided that the modulation \eqref{eq:logos_modulation} contains the DCT basis, we will uses the inverse DCT over $r[n]$ to recover its spectral components, i.e., the inverse DCT corresponds to a bank of filters at the $N$ frequencies. Notice that this is the first modulation for AirComp that is purely a detection problem. This is, the amplitude of the received signal determines the performance but plays no role in the estimation.
Notice that a linear system of equations could be built on top of the frequency components of \eqref{eq:FM_AirComp} to fully recover $m_1$ and $m_2$. However, this does not follow the philosophy of AirComp and, generally, task-oriented communications, which is recovering the function of the measurements, not the individual measurements.

Figure \ref{fig:logos_demodulation} shows the inverse DCT applied over $r[n]$ for $K=\{2,3\}$ transmitters. The modulated measurements are $\mathbf{m}=[40,60]$ and $\mathbf{m}=[10,35,55]$, respectively. Besides all the intermodulation products, in both cases the maximum frequency is located at the sum, namely,
$\Sigma=\sum_{k=1}^{K}m_k=100$.
In the presence of noise, this detection problem requires thresholding $r[n]$ to remove the noise. In the following we will characterize the noise and define an appropriate thresholding.

\begin{figure}[t]
\centering
\includegraphics[width=\columnwidth]{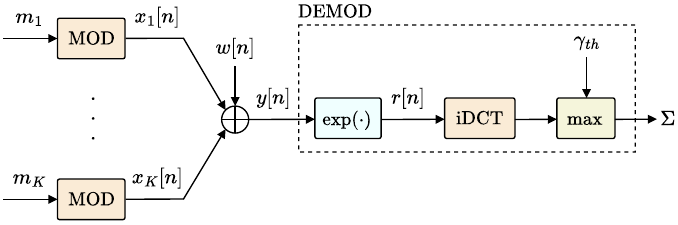}
\caption{Transmitter and receiver diagrams for implementing Log-FSK over a wireless network with $K$ transmitters and an AWGN channel.}
\label{fig:Log_FSK_comms_blocks}
\vspace{-0 pt}
\end{figure}

Figure \ref{fig:Log_FSK_comms_blocks} shows the overall AirComp system built with Log-FSK in a multiple access scheme of $K$ users and an AWGN channel. As mentioned previously, while the demodulation procedure is described in this section, the design of the modulation block may take different implementations, as the one we propose in Sec. \ref{sec:waveform_implementation}. Parameter $\gamma_{th}$ is a design parameter that determines the error probability and the required SNR.

\section{Communication Channel}
\label{sec:Communication}

Consider that each user controls the transmitted power using a scaling factor $A_k\in\mathbb{C}$ as
\begin{align}
    x_k[n] =A_k\log\left(\beta\cos_{m_k}[n]+1\right),\quad k=1,\dots,K
    \label{eq:logos_modulation_amplitude}
\end{align}
For the sake of notation simplicity and without loss of generality we constrain the analysis to $K=2$ transmitters. The signal at the input of the receiver is
\begin{equation}
    y[n] = h_1x_1[n] + h_2x_2[n] + w[n],
    \label{eq:signal_rx}
\end{equation}
\rev{where $h_k\in\mathbb{C}$ is a block flat-fading channel between transmitter $k$ and the receiver, assumed to be constant during each transmitted symbol}, and $w$ is distributed as $\mathcal{N}(0,\sigma_w^2)$ and $\sigma_w^2$ is the noise power.
Before applying the exponential postprocessing the receiver scales the received signal with $A_r\in\mathbb{C}$. The resulting signal is
\rev{
\begin{align}
    r[n] &=\exp{\left(y[n]/A_r\right)}\nonumber\\
    &=\left(\beta\cos_{m_1}[n]+1\right)^{\frac{A_1h_1}{A_r}}
    \left(\beta\cos_{m_2}[n]+1\right)^{\frac{A_2h_2}{A_r}}z[n],
    \label{eq:exp_transf}
\end{align}}
where $z[n]=\exp{(w[n]/|A_r|)}$ follows a log-normal distribution, $z[n]\sim\log\mathcal{N}(0,\sigma_w^2/|A_r|^2)$ with the following mean and variance:
\begin{align}
    \mu_z&=\exp(\sigma_w^2/2|A_r|^2)\\
    \sigma^2_z&=\exp(\sigma_w^2/|A_r|^2)\left(\exp(\sigma_w^2/|A_r|^2)-1\right)
\end{align}
\rev{
In Log-FSK, the transmitter and receiver scaling factors need to compensate the channel as $A_kh_k/A_r=1$. Otherwise, the product of cosines creates intermodulation products and the information cannot be recovered. In the rest of the paper we assume that the channel is compensated and in Section \ref{sec:power_control} we will optimize the scaling factors for maximal performance. Thus, for $A_kh_k/A_r=1$, the received signal turns into
\begin{align}
    r[n] &=
    \Biggl[\frac{\beta^2}{2}\sqrt{\frac{2}{N}}\left(\cos_{m_1+m_2}[n] + \cos_{|m_1-m_2|}[n]\right)\nonumber\\
    &\,\quad
    +\beta\left(\cos_{m_1}[n]+
    \cos_{m_2}[n]\right) +1\Biggr]z[n],
    \label{eq:FM_AirComp_noise}
\end{align}
}

The multiplicative channel designed for Log-FSK does not come for free, as the noise is now multiplicative and not normally distributed. A standard technique to deal with multiplicative noise is homomorphic filtering, in which the signal is transformed into a domain where the noise becomes additive and a linear filter can be applied \cite{Opp04}. This corresponds to applying a logarithmic transformation over $r[n]$, which converts the product into a sum; however, this procedure cannot be followed in this scenario because Log-FSK requires a multiplicative channel. Thus, Log-FSK has to deal with multiplicative noise, and requires high signal-to-noise ratio (SNR).

\subsection{\rev{Signal-to-noise ratio}}
\subsubsection{Received \textup{SNR} $\textup{(SNR}_R$\textup{)}}
The SNR at the input of the receiver (i.e., predetection SNR) and corresponding to each transmitter is
\rev{
\begin{equation}
    \text{SNR}_R= \frac{P_{\log}|A_k|^2|h_k|^2}{\sigma_w^2},
\end{equation}
where $P_{\log}$ is the power of \eqref{eq:logos_modulation}}. While the signal power increases linearly with the number of users, we define the SNR with respect to each transmitter for comparison purposes.

\subsubsection{Destination \textup{SNR} $\textup{(SNR}_\Sigma$\textup{)}}
In order to find the SNR at the destination, we particularly focus on the SNR at the frequency of interest, $\Sigma$, as the noise is multiplicative and depends on the frequency index.
While the transmitted power per user is constant, the power associated to the maximum frequency changes with the number of transmitters $K$. By inspection it can be shown that the amplitude in the absence of noise is
\begin{equation}
    A_{\Sigma}=
    \beta^K
    \left(\frac{1}{2}\right)^{K-1}
    \left(\sqrt{\frac{2}{N}}\right)^{K-1}=
    \sqrt{2N}\left(
    \sqrt{\frac{\beta^2}{2N}}\right)^K
    \label{eq:Power_sum}
\end{equation}

In the first equality, the first term corresponds to a gain provided by $\beta$ at each transmitter; the second term arises due to the product of cosines (in a product of cosines, the component associated to the sum of frequencies concentrates half the power); the last term is associated to the normalization required to compute the inverse DCT.
\rev{In the second equality, as $\beta^2\leq2N$, the maximum frequency vanishes with an increasing number of samples $N$.} 

Regarding the noise, consider \eqref{eq:exp_transf} in the frequency domain:
\begin{equation}
    \mathbf{d}=\text{DCT}\{\mathbf{r}\}=\mathbf{Qr}=\mathbf{QZp},
    \label{eq:signal_dct_vector}
\end{equation}
where we define $\mathbf{Q}\in\mathbb{R}^{N\times N}$ as the DCT matrix, $\mathbf{Z}=diag(\mathbf{z})\in\mathbb{R}^{N\times N}$ and $\mathbf{p}=\exp\left(\mathbf{x}_1/A_r+\mathbf{x}_2/A_r\right)$ with $\mathbf{x}_k\in\mathbb{R}^{N}$ being the vectorized version of \eqref{eq:logos_modulation_amplitude}. Notice that the signal in the frequency domain does not correspond to multiplicative noise over the noiseless signal in the frequency domain; conversely, the noise can be seen as a perturbation of matrix $\mathbf{Q}$. To study the noise distribution we express \eqref{eq:signal_dct_vector} as signal plus noise:
\begin{equation}
    \mathbf{d}=\mathbf{s}+\bm{\varepsilon}=\mathbf{Qp} + \mathbf{Q(Z-I)p},
    \label{eq:signal_dct_vector_additive}
\end{equation}
\rev{where $\mathbf{s}$ is the noiseless signal and $\varepsilon$ is the noise.} The second term reflects that the noise term is correlated with the transmitted signal. Each entry of the noise in \eqref{eq:signal_dct_vector_additive} corresponds to
\begin{align}
    \varepsilon[\ell]=\sum_{n=0}^{N-1}\mathbf{p}[n]\mathbf{q}_\ell[n]\left(z[n]-1\right),
    \label{eq:noise_distribution}
\end{align}
where $\mathbf{q}_\ell$ is the $\ell$th DCT basis. Each of the terms in the sum corresponds to a weighted shifted log-normal distributed as
\begin{align}
    \log\mathcal{N}\left(\log(\mathbf{p}[n]\mathbf{q}_\ell[n]), \frac{\sigma_w^2}{|A_r|^2},-\mathbf{p}[n]\mathbf{q}_\ell[n]\right),
\end{align}
where the last term corresponds to the shift. There is existing literature on the probability distribution of a sum of log-normal random variables. In \cite{Cobb12} the authors propose to use the Fenton-Wilkinson method to estimate the parameters for a single log-normal distribution that approximates the sum of log-normals; in \cite{Wu05} they use the short Gauss-Hermite approximation of the moment generating function, while \cite{Nag02} uses an extended convolution. To the best of our knowledge, \cite{Ras02} is the only work on weighted sums of log-normals. Nevertheless, all the previous results rely on numerical integration methods. To the extend of our knowledge, there is no closed-form expression for the distribution of \eqref{eq:noise_distribution}.

We observe that, in practice, the central limit theorem applies and we can approximate $\bm{\varepsilon}[\ell]\sim \mathcal{N}(\mu_{\varepsilon[\ell]}, \sigma^2_{\varepsilon[\ell]})$. See Appendix \ref{noise_distribution} for several graphical representations of the noise distribution. Notice that these parameters depend on the index $\ell$ as the noise is multiplicative. We provide closed-form expressions for the mean and variance in Theorem \ref{thm: mean_var}.

\begin{theorem}
\label{thm: mean_var}
The mean and variance of the multiplicative noise at the frequency index $\ell=\Sigma$ are
\begin{align}
  \mu_{\varepsilon} = A_\Sigma(\mu_z-1)\qquad 
  \sigma^2_{\varepsilon} = \sigma_z^2P_p
  \label{eq:noise_demod_var}
\end{align}
where $P_p$ is the power of noiseless signal in \eqref{eq:FM_AirComp_noise}, this is,
\rev{
\begin{equation}
    P_p = \frac{1}{N}\sum_{n=0}^{N-1} \exp\left(\frac{2}{A_r}\sum_{k=1}^{K}x_k[n]\right)
    \approx (\beta^2+1)^K
\end{equation}}

\begin{proof}
The proofs can be found in Appendix \ref{proofs_noise_demod}.
\end{proof}
\end{theorem}

\begin{corollary}
While $P_p$ is fixed for a given $K$, it increases exponentially with $K$, which limits the applicability of Log-FSK to non-massive scenarios.
\label{cor:non_massive}
\end{corollary}

\rev{
Since the mean of the noise, $\mu_\varepsilon$, is known at the receiver side, this bias can be subtracted from the received signal to increase the SNR.
}
Thus, the SNR at frequency $\Sigma$ can be expressed as
\rev{
\begin{align}
    \text{SNR}_\Sigma=\frac{A_\Sigma^2}{\sigma_\varepsilon^2}
    =\frac{\beta^{2K}}{(2N)^{K-1}(\beta^2+1)^K}
    \frac{\exp(-\sigma_w^2/|A_r|^2)}{\exp(\sigma_w^2/|A_r|^2)-1}
    \label{eq:snr_destination}
\end{align}
}

\begin{remark}
Notice that the SNR definition in \eqref{eq:snr_destination} differs from the usual, where the denominator contains the signals uncorrelated with the signal of interest. 
\end{remark}

Finally, the expression of \eqref{eq:snr_destination} in high-SNR ($\text{SNR}_R$) regime is provided in Theorem \ref{thm:high_snr}.

\begin{theorem}
\label{thm:high_snr}
In high-SNR regime, the SNR at destination can be approximated as
\upshape
\rev{
\begin{equation}
    \text{SNR}_\Sigma \approx
    \frac{\beta^{2K}}{(2N)^{K-1}(\beta^2+1)^K}
    \left(
    \frac{|A_r|^2}{\sigma_w^2}
    \right)^2
    \label{eq:high_SNR}
\end{equation}}
\begin{proof}
\rev{
The proof is based on the first-order Taylor approximation of the exponential function. The full proof is omitted for brevity.}
\end{proof}
\end{theorem}

\rev{
As stated in Corollary \ref{cor:non_massive}, the noise power increases with $K$. Furthermore, the effective $\text{SNR}_\Sigma$ decreases with $K$. This combined effect leads to a reduction in the effective SNR, which limits the applicability of Log-FSK in massive scenarios.
}



\rev{
\subsection{Error probability}
Log-FSK requires the detection of the maximum frequency, which translates into a classical detection problem in digital modulations. Unlike classical $M$-ary FSK, where any incorrect frequency detection is considered an error, an error occurs only if 
$\mathbf{d}$ detects a higher frequency component than $\Sigma$. Lower spectral components, $\ell<\Sigma$, do not contribute to the error, even if they have higher amplitudes because $\Sigma$ will still be detected. Consider the \revi{signal \eqref{eq:signal_dct_vector_additive} particularized at the $\ell$-th frequency component},
\begin{equation}
    \mathbf{d}[\ell]=
    \begin{cases}
      A_{\Sigma}+\varepsilon[\ell] & \ell=\Sigma\\
      \varepsilon[\ell] & \ell>\Sigma
    \end{cases}
\end{equation}
where $\varepsilon[\ell]\sim\mathcal{N}(0,\sigma_\varepsilon^2)$. The error probability produced by Log-FSK is presented in Theorem \ref{thm:error_probability}. Unlike classical $M$-ary FSK, where all symbols have an equal likelihood of error, Log-FSK inherently reduces the number of competing frequency components that can cause errors. This results in a non-uniform error distribution, where certain frequency bands exhibit improved detection performance. 
\begin{theorem}
\label{thm:error_probability}
The error probability of Log-FSK can be approximated as
\begin{equation}
    P_e \approx
    \left(N-\Sigma\right)
    Q\left(\sqrt{\text{SNR}_{\Sigma}}\right)\label{eq:error_probability}
\end{equation}
\begin{proof}
\rev{
The proof can be found in the Appendix \ref{proof_error_prbability}.}
\end{proof}
\end{theorem}
}

\begin{figure}[t]
\centering
\includegraphics[width=\columnwidth]{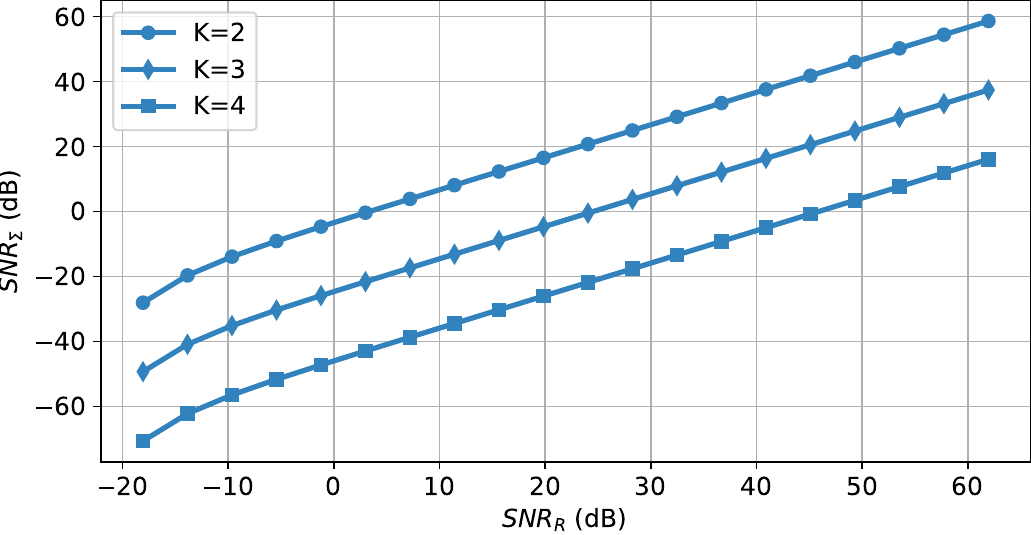}
\caption{\rev{SNR at destination ($\text{SNR}_\Sigma$) with respect to received SNR ($\text{SNR}_R$) for different $K$.}}
\label{fig:SNR_R_SNR_D}
\vspace{-0 pt}
\end{figure}

Figure \ref{fig:SNR_R_SNR_D} shows the relationship between $\text{SNR}_R$ and $\text{SNR}_\Sigma$ for different number of samples $N$ and users $K$. Considering a fixed noise power, Figure \ref{fig:SNR_R_SNR_D} shows the power $\text{SNR}_\Sigma$ with respect to the transmitted power. Below $SNR_R=20$ dB, Log-FSK exhibits a threshold effect under which $\text{SNR}_\Sigma$ drops rapidly. This is a natural behavior in any frequency modulated system, and a minimum $\text{SNR}_R$ is required to work above the threshold. 
The procedure to determine this threshold is the following: The performance of Log-FSK is determined by the correct detection of the peak at frequency $\Sigma$. 
While $SNR_{\Sigma}$ degrades with $K$, Log-FSK can still be used beyond $K=2$. For $K=3 \text{ and } 4$ it requires around 30 and 50 dB in transmission, respectively. 
Moreover, there are techniques to extend the threshold, such as phase locked loops that can improve the performance of Log-FSK at low SNR.

\rev{
\subsection{Mean Squared Error}
The usual metric of an AirComp system is not the SNR, but the MSE. In this respect, we decompose the MSE into contributions from all possible frequencies detected:
\begin{equation}
    \text{MSE} = \sum_{\ell=0}^{N-1} 
    Pr(\ell|\Sigma)(\ell-\Sigma)^2,
\end{equation}
where $Pr(\ell|\Sigma)$ is the probability that frequency $\ell$ is detected as the maximum when frequency $\Sigma$ is the ground truth maximum. As in the error probability, notice that only the $\ell>\Sigma$ components incur in errors. The final closed-form expression of the MSE is found in Theorem \ref{thm:mse}.
\begin{theorem}
\label{thm:mse}
The MSE produced by Log-FSK can be approximated by
\begin{equation}
    \text{MSE}\approx
    \frac{
    \left(N-\Sigma\right)
    \left(N-\Sigma-1\right)
    \left(2(N-\Sigma)-1\right)}{6}
    Q\left(\sqrt{\text{SNR}_{\Sigma}}\right)
    \label{eq:mse}
\end{equation}
\begin{proof}
\rev{
The proof can be found in the Appendix \ref{thm:mse_proof}.}
\end{proof}
\end{theorem}
The derived MSE expression shows that MSE is directly proportional to the error probability $P_e$, meaning that reducing detection errors leads to lower MSE. Additionally, MSE grows with the number of possible incorrect detections, particularly when higher frequency components are mistakenly identified. In general, maximizing the use of high-frequency symbols is beneficial, as it reduces the likelihood of incorrect higher-frequency detections, thereby lowering MSE. However, keeping $N$ small is also crucial, as a larger number of samples increases the range of possible incorrect detections. This creates a trade-off between quantization error and MSE: a finer frequency resolution (i.e., large $N$) reduces quantization error but increases MSE due to more possible errors.
}

\rev{
\subsection{Power control}
\label{sec:power_control}
In classical communication systems the goal is to minimize the error probability. Conversely, in AirComp the goal is to minimize the MSE. Since the MSE is proportional to $P_e$, in Log-FSK both problems are equivalent.
In the following we formulate an optimization problem to adapt the power control in transmission and reception, as well as the design parameter $\beta$, to minimize the MSE.
\begin{subequations}
\begin{align}
\nonumber
& \underset{\beta,\{A_k\}_{k=1}^{K}, A_r}{\text{minimize}} ~~~~
\eqref{eq:mse} \\
& \hspace{-14pt}\qquad\text{subject to}~~~~~~
A_kh_k=A_r,~~~~k=1,\dots,K
\label{const:channel}\\
& ~~~~~~~~~~~~~~~~~~~
|A_k|^2\leq P_k,~~~~k=1,\dots,K\label{const:power_tx}\\
& ~~~~~~~~~~~~~~~~~~~
|A_r|^2\leq P_R
\label{const:power_rx}
\end{align}
\label{eq:min_mse}
\end{subequations}
where constraint \eqref{const:channel} ensures channel compensation, \eqref{const:power_tx} limits the transmission power at each transmitter and \eqref{const:power_rx} at the receiver side.
\revi{We note that $P_R$ has no direct physical meaning, as it acts purely as a scaling factor, but it serves as a useful design parameter to balance MSE and overall system power.} 
Channel compensation must account for both magnitude and phase to ensure proper signal alignment at the receiver. Since the receiver only has a single combined term for all transmitters, phase compensation is assumed to be handled individually at each transmitter. Therefore, we assume perfect \revi{phase synchronization with respect to the carrier frequency} and focus solely on adjusting the magnitude of the channel.
}

\rev{
Minimizing the MSE in \eqref{eq:min_mse} is directly linked to minimizing the error probability $P_e$, as both are proportional. Since $P_e$ is driven by the $Q$-function, which depends on the $\text{SNR}_{\Sigma}$, reducing the MSE is equivalent to maximizing the $\text{SNR}_{\Sigma}$. Unlike traditional AirComp designs, Log-FSK follows the traditional design principle in communications of maximizing SNR. Nonetheless, this problem is not convex because constraint \eqref{const:channel} is not convex due to the coupling between the variables. We reformulate the previous problem to make it convex as
\begin{subequations}
\begin{align}
\nonumber
& \underset{\beta, A_r}{\text{maximize}} ~~~~
\eqref{eq:snr_destination} \\
& \hspace{-20pt}\qquad\text{subject to}~~~~
|A_r|^2\leq P_R
\label{const:channel_v2}
\end{align}
\label{eq:max_snr}
\end{subequations}
}

It is easy to show that the new cost function is convex in $\beta$ and $A_r$. Thus, the minimum MSE and the maximum $\text{SNR}_{\Sigma}$ is achieved at maximum $\beta$ and $A_r$. \revi{Since $\beta<\sqrt{N/2}$, we choose $\beta=0.99\sqrt{N/2}$ and $|A_r|=\sqrt{P_R}$. Then, each user adapts the power control as $|A_k|=|A_r|/|h_k|$, as long as $|A_k|\leq\sqrt{P_k}$; otherwise they remain silent. In conclusion, each transmitter performs power control as
\begin{equation}
    |A_k| = \begin{cases}  
\min\left(\frac{\sqrt{P_R}}{|h_k|}, \sqrt{P_k}\right), & \text{if } \frac{P_R}{|h_k|^2} \leq P_k, \\  
0, & \text{otherwise}
\end{cases}
\label{eq:power_control}
\end{equation}}

\rev{
The optimal power control for Log-FSK follows a threshold-based structure, where a device can only transmit if its channel gain is sufficiently large to satisfy the power constraint. This leads to a natural cutoff effect, where only devices with favorable channel conditions can participate in transmission.}
\revi{
Although dropping users may bias the AirComp result, this structure ensures efficient power usage, as devices with stronger channels transmit optimally while those in deep fade remain silent, preventing excessive power consumption.
}

\section{Computation of functions}
\rev{
This section explores functions computable with Log-FSK, starting with scenarios for its application in low-connectivity networks. We then cover functions like subtraction and maximum, and expand the possibilities using the functional nomographic representation theorem.
}

\subsection{Scenarios for Log-FSK}
\rev{
Log-FSK is not designed for massive access, but it suits scenarios with low network connectivity. For example, \cite{Guirado23} implements an AirComp scheme with only three transmitters for intra-chip computations. In wireless sensor networks (WSN), communication is often limited to pairs of nodes, as in physical layer network coding \cite{zhang2006networkcoding}. Linear and cyclical topologies follow this pattern: In vehicle platooning \cite{Lee22}, each vehicle communicates only with its immediate neighbors, while in LEO satellite constellations \cite{Leyva23}, satellites interact within a single orbital plane or with up to four neighbors.
}

\subsection{Subtraction}
In this work we frame the difference between two data points as a mechanism for simultaneous detection and localization. Consider a set of nodes taking measurements and a detection task, where the true hypothesis at node $k$ is associated to
\begin{equation}
    \mathcal{H}_1^{(k)}:
    f(m_{k-1}, m_{k+1})= |m_{k+1}-m_{k-1}|>\theta,
\end{equation}
while $\mathcal{H}_0$ happens otherwise. We assume a linear topology where node $k$ can communicate with its preceding and succeeding nodes, this is, $k-1$ and $k+1$, respectively.
As shown in \eqref{eq:FM_AirComp}, there is also a term associated to the difference or subtraction of measurements (see also Figure \ref{fig:logos_demodulation} for $K=2$). Thus, to compute this function the system requires no modifications besides taking the other tone with minimum amplitude that is not the maximum frequency. Since this term has the same power as the maximum frequency, it offers the same capabilities as the sum function. In other words, computing the addition or the difference using Log-FSK has no difference in terms of communication and performance. 


\subsection{Maximum}
Computing the maximum is a crucial statistic for distributed data. It can be used for monitoring, consensus and detection tasks, among others.
When possible, the philosophy in AirComp is to decompose the function with preprocessing and postprocessing functions (see Sec. \ref{sec:nomographic}). In the case of the maximum function, it can be approximated using the $p$-norm with $p\rightarrow\infty$ \cite{GOldenbaum13}, or using the log-sum-exp:
\begin{equation}
    \max(m_k,m_l)\approx
    \log\left(\exp(m_k)+\exp(m_l)\right)
    \label{eq:max_nomographic}
\end{equation}

However, this may exceed the dynamic range due to the exponential transformation. What we propose is using
\begin{equation}
    \max(m_k,m_l)=
    \frac{m_k+m_l+|m_k-m_l|}{2},
\end{equation}
which is exact for two measurements. Notice that this expression matches once again with the frequencies present in \eqref{eq:FM_AirComp}. Thus, to implement the maximum function the receiver needs to threshold the signal, remove the maximum frequency and the DC, and sum all the three frequency indices.


\subsection{Nomographic functions}
\label{sec:nomographic}
The Log-FSK modulation is versatile because it allows to compute several functions simultaneously in a single transmission. To compute more complex functions, one can use the nomographic decomposition shown in \eqref{eq: nomographic}.
As discussed in Sec. \ref{sec:modulation}, this was the reasoning behind the inception of the Log-FSK waveform for frequency-based AirComp. However, this nomographic function representation can also be applied to the modulated variable to compute other functions. For instance, to compute the product of measurements each transmitter modulates $\varphi(m_k)=\log(m_k)$ using Log-FSK. The receiver always extracts the maximum frequency, since it corresponds to the sum in \eqref{eq: nomographic}. Finally, the receiver applies the exponential over the frequency index, which yields
\begin{equation}
    \exp\left(\sum_{k=1}^K \log(m_k)\right)=
    \exp\left(\log\left(\prod_{k=1}^K m_k\right)\right)=
    \prod_{k=1}^K m_k,
    \label{eq:product_nomographic}
\end{equation}
where the argument of the exponential corresponds to $\Sigma$.
We refer the reader to \cite{sahin22, GOldenbaum13} for an extensive list of functions and its nomographic representation.


One challenge of using nomographic representations with linear AirComp is that it affects the power consumption. For instance, the approximation for the maximum function requires an exponential as a pre-processing, which increases the transmitted power exponentially. Conversely, the transmitted power by Log-FSK is independent of the modulated information. Note, however, that these transformations may require an increase in the transmission bandwidth.

\section{Numerical results}
In the following we will evaluate the performance of Log-FSK in terms of MSE and power consumption with respect to linear analog AirComp.
\rev{
The linear analog modulation is implemented with DSB, i.e.,
\begin{equation}
    x_k[n] = \bar{A}_km_k\cos_{m_0}[n],
    \label{eq:dsb_mod}
\end{equation}
where the information $m_k$ is encoded in the amplitude of the carrier and $m_0$ is a default carrier frequency. For consistency with Log-FSK, $A_k$ is used to adjust the transmitted power and combat the fading.}

\subsection{MSE performance}
\begin{figure}[t]
\centering
\includegraphics[width=\columnwidth]{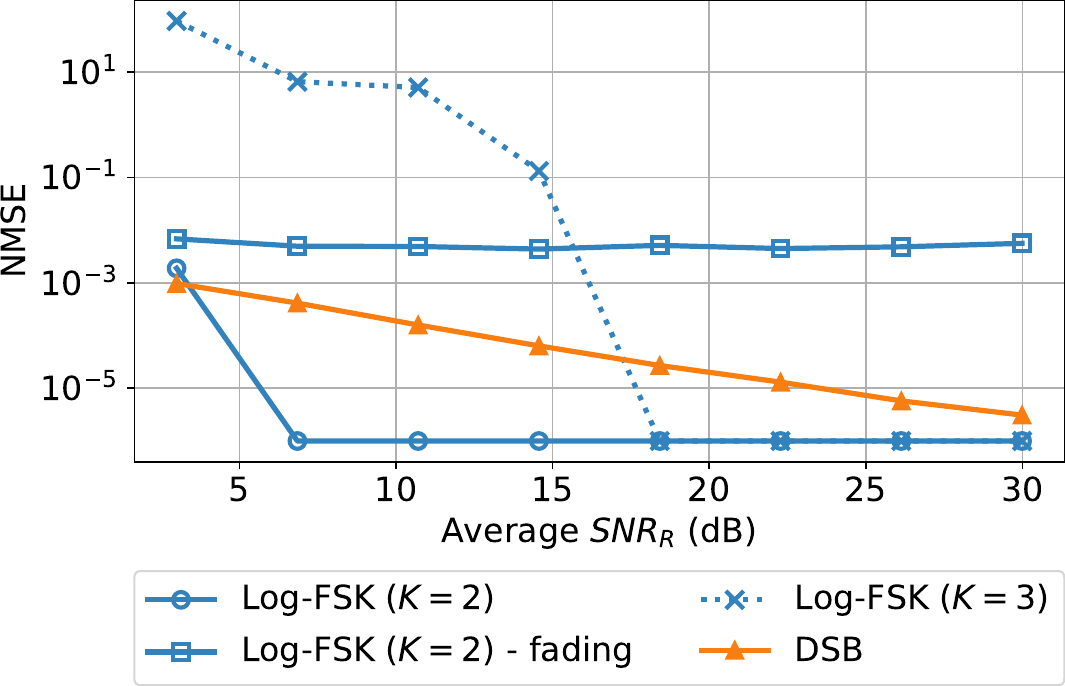}
\caption{NMSE with respect to average $SNR_R$ for Log-FSK and linear AirComp implemented with DSB at different $K$ \revi{and channel conditions}.}
\label{fig:MSE_SNR_R}
\vspace{-0 pt}
\end{figure}

\revi{Figure \ref{fig:MSE_SNR_R} shows the performance of Log-FSK for $K=2$ for the average function. We also compare it against the benchmark DSB, and evaluate Log-FSK under increased number of transmitters ($K=3$) and under Rayleigh fading conditions.
In the latter, the in-phase and quadrature components of the channel follow a 
$\mathcal{N}(0,1)$ distribution, and the magnitude of the channel $|h_k|$ follows a Rayleigh distribution.} In both Log-FSK and DSB the measurements $m_k$ come from a uniform distribution in $[0,N/2]$ \revi{with $N=256$} and the NMSE is averaged over $10^4$ realizations. Since the $\text{SNR}_R$ in DSB depends on the transmitted information, we propose to evaluate both schemes in terms of average $\text{SNR}_R$, this is, at the same total transmitted power:
\begin{equation}
    \text{Average } \text{SNR}_R=\frac{1}{K} \sum_{k=1}^K \text{SNR}_R=\frac{1}{K\sigma_w^2} \sum_{k=1}^K \bar{A}_{k}^2m_k^2
    \label{eq:average_snr}
\end{equation}

\rev{
As expected, Log-FSK achieves zero MSE when working above the threshold SNR, which is around $7$ dB for $K=2$ and $19$ dB for $K=3$. The performance of DSB is almost the same for both cases. Log-FSK offers a significant advantage when operating above the threshold, where it achieves zero MSE, making it completely error-free. While DSB may perform better in extremely low-SNR conditions due to its linear noise impact, Log-FSK ensures perfect accuracy once the threshold is exceeded, providing a clear performance advantage in practical scenarios where sufficient SNR is available.}
\revi{Fading degrades performance, primarily due to bias introduced by dropped users rather than noise, highlighting the need for clear line-of-sight conditions for Log-FSK.}

\subsection{Power consumption}
In linear analog AirComp it is not possible to adjust the transmission to a specific SNR level. This is because \eqref{eq:dsb_mod} encodes the information in the amplitude and modifying the transmitted power alters the encoded data. Thus, to adjust the power the system needs to know the average transmitted power as \eqref{eq:average_snr} indicates. This is not possible, as it requires centralizing the data, which is the ultimate goal of AirComp systems. Another drawback of linear analog modulations is that each transmitter experiences a different power consumption, which can drain the battery of tightly constrained devices.
Conversely, Log-FSK is a pure frequency modulation and the corresponding transmitted power is independent of the modulated information. In this respect, adjusting the SNR can be performed in a distributed fashion.

As mentioned in \ref{sec:nomographic}, in the case of computing functions using a nomographic representation, the transmitted power depends on the initial measurement distribution, as well as on the preprocessing performed at every transmitter.
We evaluate the performance of both modulations for a uniform and a Gaussian distribution over $m_k$ and for three different functions:
\begin{itemize}
    \item Sum: this is standard AirComp, which does not require any preprocessing or postprocessing function.
    \item Product: this corresponds to \eqref{eq:product_nomographic}, with the logarithm as the preprocessing function.
    \item Maximum: this corresponds to \eqref{eq:max_nomographic}, with the exponential as the preprocessing function.
\end{itemize}

\begin{figure}[t]
\centering
\includegraphics[width=\columnwidth]{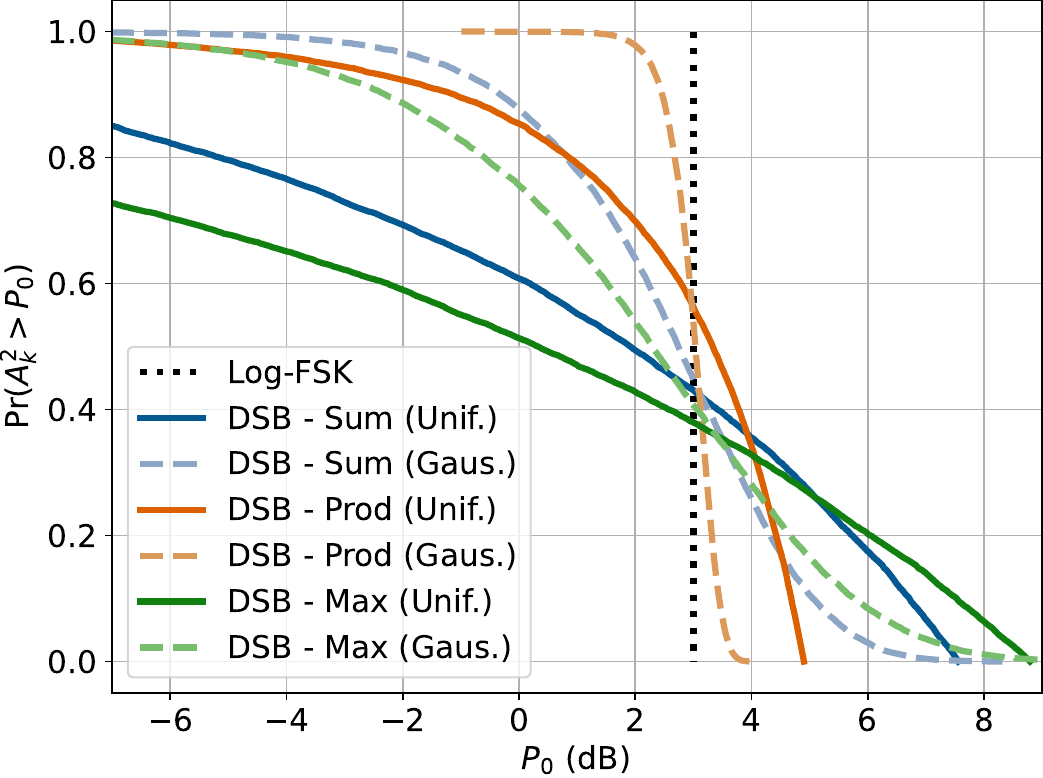}
\caption{\rev{Transmitted power distribution to achieve an average $\text{SNR}_R=10$ dB with Log-FSK and DSB under fading channels. The scenario considers \revi{$N=256$}, $K=2$ users, two data distributions on $m_k$ (uniform and Gaussian) and three different functions (sum, product and maximum).}}
\label{fig:cdf_tx_power}
\vspace{-0 pt}
\end{figure}

Figure \ref{fig:cdf_tx_power} shows the transmitted power distribution (cumulative distribution function) per user for Log-FSK and DSB under the same previous fading channels.
\revi{We consider $N=256$,} $K=2$ transmitters and $\text{SNR}_R=10$ dB, with the power control in \eqref{eq:power_control} applied to optimize transmission power.
As expected, the the transmitted power by Log-FSK is independent of the power distribution and preprocessing. While in DSB approximately half of the users consume less power than in Log-FSK, which happens at the expenses of the rest transmitting more power $P_o$, and some users transmitting at high power, with the associated risk of draining. Furthermore, the power distribution in DSB highly depends on the distribution of data and function to be computed, which does not happen for Log-FSK.

Moreover, the peak-to-average-power ratio (PAPR) can be reduced by up to $1.65$ dB compared to standard cosine waveforms (e.g., DSB) improving power efficiency and reducing nonlinear distortion in practical implementations.

\section{Concluding Remarks}
In this paper we propose Log-FSK, a novel frequency modulation for AirComp that implements DA as a multiple access. The non-linear waveform design allows to recover a frequency component that corresponds to the sum of the individual transmitted frequencies.
We provide a demodulation procedure that relies on a bank of filters (e.g., DCT) and extracting the maximum frequency. Furthermore, as the scheme corresponds to detection problem, the amplitude of the received signal provides no information for the estimation. This is, Log-FSK inherits the communication benefits of angular modulations.

\rev{We provide the analytical performance of Log-FSK in terms of MSE in an AWGN and fading channels.} 
Log-FSK performs reliably in scenarios with up to 4 users, which aligns with many practical applications requiring low-to-moderate user density.
Then, we show how the modulation can integrate nomographic function representation to compute functions beyond the sum and not affecting the transmitted power. 
We present  numerical results showcasing the performance of Log-FSK in terms of MSE and energy consumption with respect to amplitude modulations.

\rev{
Log-FSK advances AirComp modulation design by integrating computation within communication, reducing processing complexity at both the transmitter and receiver while lowering overall latency.
Although its non-linear processing introduces challenges in managing multiplicative noise, it also presents opportunities for innovative signal design.} Future work may explore its potential for physical-layer network coding, particularly in scenarios with two simultaneous transmitters.

\appendix
\vspace{-15pt}
\rev{
\subsection{Proof of Lemma \ref{thm:mean}}
\label{app:proof_lemma_mean}
\revi{
Under the continuous-time approximation:
\begin{align}
    \bar{x}_k&=\frac{1}{2\pi}\int_{0}^{2\pi}
    \log\left(\alpha\cos(x)+1 \right)dx\nonumber\\
    &\overset{(a)}{=}\frac{1}{2\pi}\int_{0}^{2\pi}
    \sum_{i=1}^\infty
    (-1)^{i+1}\frac{\alpha^i}{i}\cos^i(x)dx\nonumber\\
    &=\sum_{i=1}^\infty
    (-1)^{i+1}\frac{\alpha^i}{i}\frac{1}{2\pi}
    \int_{0}^{2\pi}
    \cos^i(x)dx\nonumber\\
    &\overset{(b)}{=}\sum_{j=1}^\infty
    (-1)^{2j+1}\frac{\alpha^2j}{2j}
    \frac{1}{2^{2j}}{2m\choose m}\nonumber\\
    &\overset{(c)}{=}\log\left(
    \frac{1+\sqrt{1-\alpha^2}}{2}
    \right)
\end{align}
In (a) we substitute the Taylor series expansion of the logarithm. In (b) only the even coefficients are different from zero. Equivalence (c) corresponds to the complete elliptic integral of the first kind. The final result is obtained by setting $\alpha=\beta\sqrt{2/N}$.}
\hfill$\square$}

\subsection{Proof of Theorem \ref{thm:spectrum}}
\label{app:proof_thm_spectrum}
\rev{
The complex cepstrum associated to a signal $x[n]$ corresponds to a sequence $\hat{x}[n]$, whose $z$-transform is $\hat{X}(z)=\log(X(z))$. Thus, we need to work with the signal in \eqref{eq:logos_modulation} as a signal in the frequency domain (i.e., understand $n$ as frequency).}
\rev{
Substituting the cosine in \eqref{eq:logos_modulation} by complex exponentials results in
\begin{equation}
    X(z)=\beta\sqrt{\frac{2}{N}}\frac{z^{-m}+z^{m}}{2} + 1,
\end{equation}}
\rev{
where $z$ is a complex exponential \revi{$m_k=m$}. Following the same notation as in the cepstrum context, the complex cepstrum in the frequency domain is $\hat{X}(z)=\log X(z)$. Using the power series representation of the logarithm, this can be rewritten as
\begin{equation}
    \hat{X}(z)=-\sum_{i=1}^{\infty}\frac{(-1)^{i}}{i}\left(
    \frac{\beta}{2}\sqrt{\frac{2}{N}}\left(z^{m}+z^{-m} \right) \right)^i,
    \label{eq:cepstrum_z_transform}
\end{equation}
whose inverse $z$-transform is the complex cepstrum and corresponds to \eqref{eq:logos_modulation} in the frequency domain:
\begin{equation}
    \hat{x}[n]=\mu_x\delta[n]-\sum_{i=1}^{\infty}\frac{(-1)^{i}}{i}\left(\frac{\beta}{2N}\right)^i
    \delta[n-im]
\end{equation}
for $n>0$, in which we have included the mean of the signal, $\mu_x$, at the origin. Finally, in the notation of spectral analysis, the result is obtained by changing the index $n$ by $l$. \hfill$\square$}

\subsection{Graphical representation of the noise distribution}
\label{noise_distribution}
\begin{figure}[t]
\centering
     \begin{subfigure}[b]{0.49\columnwidth}
         \includegraphics[width=\columnwidth]{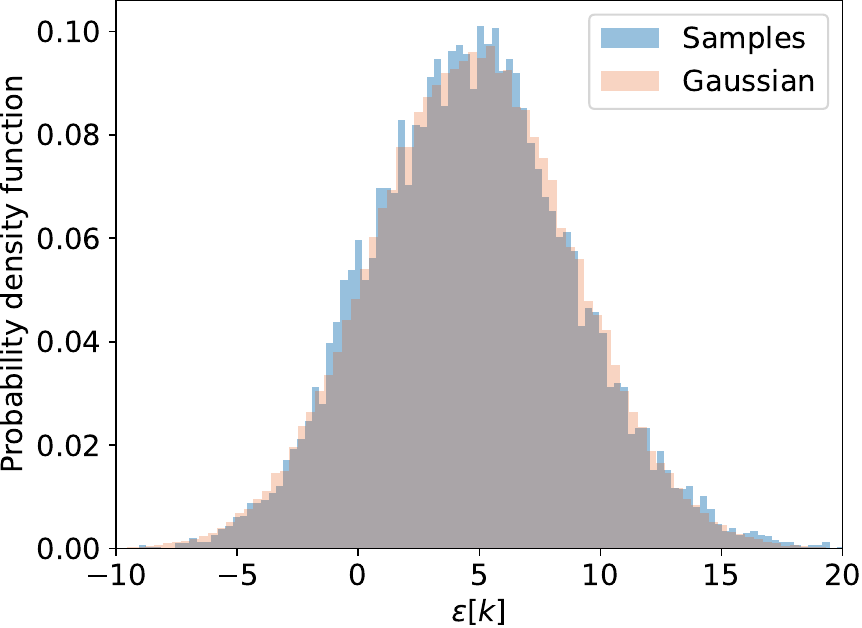}
         \caption[]{Histogram.}
         \label{fig:hist_7dB}
     \end{subfigure}
     \begin{subfigure}[b]{0.49\columnwidth}
         \includegraphics[width=\columnwidth]{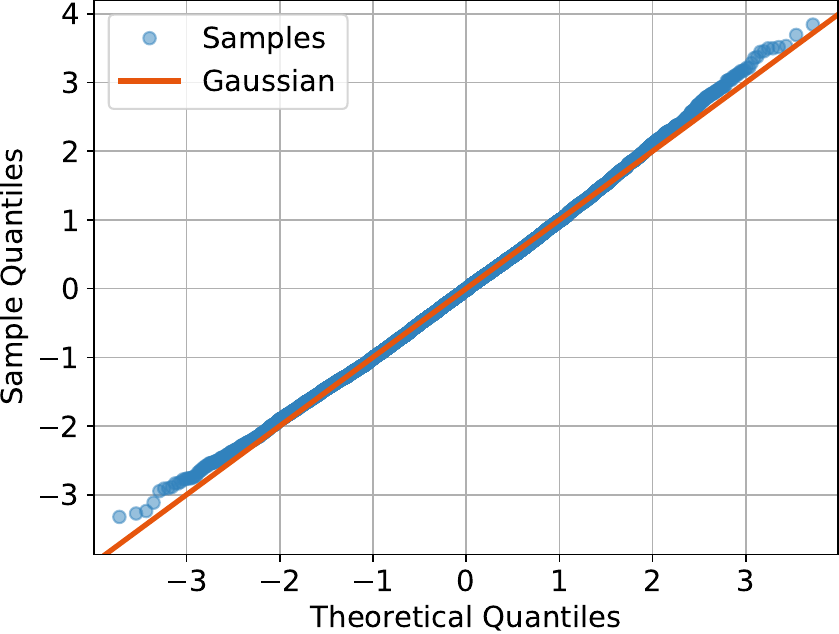}
         \caption[]{QQ-plot.}
         \label{fig:qq_7dB}
     \end{subfigure}
     \hspace{5pt}
     \newline
     \begin{subfigure}[b]{0.49\columnwidth}
         \includegraphics[width=\columnwidth]{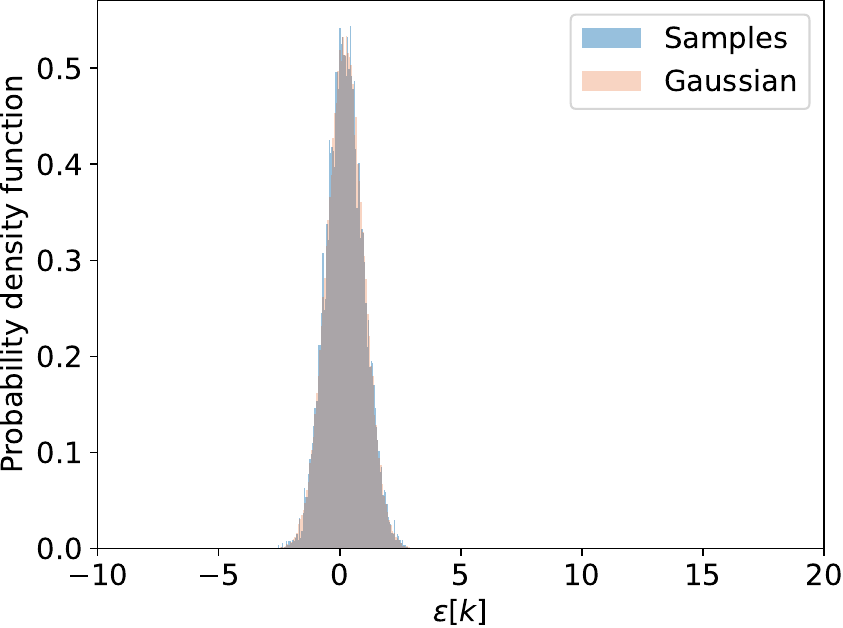}
         \caption[]{Histogram.}
         \label{fig:hist_20dB}
     \end{subfigure}
     \begin{subfigure}[b]{0.49\columnwidth}
         \includegraphics[width=\columnwidth]{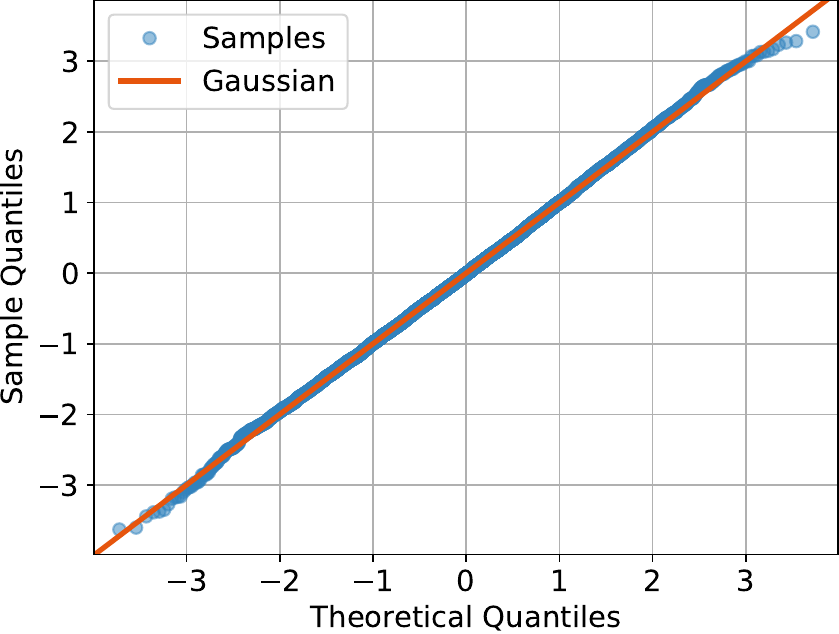}
         \caption[]{QQ-plot.}
         \label{fig:qq_20dB}
     \end{subfigure}

     \caption{Comparison of the noise sample distribution at $\ell=\Sigma$ and a synthetic Gaussian distribution with the equivalent mean and variance at (top) $\text{SNR}_R=7$ dB and (bottom) $\text{SNR}_R=20$ dB \rev{for $N=512$}.}
     \label{fig:noise_distribution}
\end{figure}

Figure \ref{fig:noise_distribution} shows the histogram and quantile–quantile plot for the noise distribution in \eqref{eq:noise_distribution} at different $\text{SNR}_R$. While this does not guarantee that the noise is Gaussian by nature, it is safe to assume so for the theoretical analysis in the range of SNR values.
These results have been generated using $10^4$ experiments in a Monte Carlo fashion.

\subsection{Proof of Theorem \ref{thm: mean_var}}
\label{proofs_noise_demod}

Consider a signal $\mathbf{p}$ defined as
\begin{equation}
    \mathbf{p}[n]=\sum_{\ell=1}^L A_\ell\cos_{m_{\ell}}[n],
    \label{eq: p_signal}
\end{equation}
which corresponds to \eqref{eq:FM_AirComp_noise} in the absence of noise for $L=5$ (the DC component corresponds to $m_{\ell}=0$). We further define $\mathcal{M}=\{m_1,\dots,m_L\}$ as the set of frequency indices where signal is present.
The $\ell$th noise sample after demodulation, i.e., the noise found at discrete frequency $\ell$ corresponds to
\begin{equation}
    \bm{\varepsilon}[\ell]= \mathbf{q}_\ell^T\mathbf{(Z-I)p}
\end{equation}
By definition, $\mathbf{p}$ contains the DCT basis, and the product with 
$\mathbf{q}_\ell$ is different from zero and equal to one when $\ell=m_l$. In the following we will find the mean and variance of $\bm{\varepsilon}[\ell]$. Firstly, the mean is computed as
\begin{align}
    \mu_{\varepsilon[\ell]}&=\mathbb{E}\{\bm{\varepsilon}[\ell]\}=
    \mathbf{q}_\ell^T\mathbb{E}\{\mathbf{Z}\}\mathbf{p} - \mathbf{q}_\ell^T\mathbf{p}\nonumber\\
    &=(\mu_z-1)\mathbf{q}_\ell^T\mathbf{p}=
    \begin{cases}
      0 & k\not\in\mathcal{M}\\
      A_{\ell}(\mu_z-1) & k\in\mathcal{M}
    \end{cases}
\end{align}

And the variance:
\begin{align}
    \sigma^2_{\varepsilon[\ell]}&=\mathbb{E}\{(\bm{\varepsilon}[\ell]-\mu_{\varepsilon[\ell]})^2\}=\mathbb{E}\{\mathbf{p}^T\mathbf{Zq_\ell q_\ell}^T\mathbf{Zp}\}\nonumber\\
    &=\mathbf{p}^T\mathbf{Q}_\ell^T\mathbb{E}\{\mathbf{zz}^T\}\mathbf{Q_\ell p}=\sigma_z^2\mathbf{p}^T\mathbf{Q}_\ell^T\mathbf{Q_\ell p}\nonumber\\
    &=\sigma_z^2\sum_{n=0}^{N-1}\sum_{l=1}^{L} A_l^2\cos_{m_{\ell}}^2[n]\cos_{\ell}^2[n]\nonumber\\
    &=\sigma_z^2\sum_{l=1}^{L} \frac{A_l^2}{N^2}\sum_{n=0}^{N-1} \biggl(
    \frac{1}{2}\cos_{m_{\ell}+\ell}[n] +
    \frac{1}{2}\cos_{m_{\ell}-\ell}[n]\nonumber\\
    &\qquad\qquad\qquad\qquad
    +\cos_{m_{\ell}}[n] +
    \cos_{\ell}[n] + 1\biggl)\nonumber\\
    &=\frac{\sigma_z^2}{N}\sum_{{\ell}=1}^{L}A_{\ell}^2S(m_{\ell},\ell)
\end{align}
where we define $\mathbf{Q}_\ell=diag(\mathbf{q}_\ell)$. The sum $S(m_{\ell},\ell)$ corresponds to the sum of cosine averages. Depending on the values of $m_{\ell}$ and $\ell$ this term has different results:
\begin{align}
    S(m_{\ell},\ell)=
    \begin{cases}
      4 & m_{\ell}=0 \text{ and } \ell=0\\
      2 & m_{\ell}=0 \text{ or } \ell=0\\
      1 & m_{\ell}\neq \ell\\
      3/2 & m_{\ell}=\ell
    \end{cases}
\end{align}

As a reference value, we take the variance where there is no correlation between the signal and the noise ($\ell\not\in\mathcal{M}$):
\begin{align}
    \sigma^2_{\varepsilon}&=
    \frac{\sigma_z^2}{N} \sum_{l=1}^{L}A_l^2=
    \frac{\sigma_z^2}{N} \sum_{k=0}^{N-1}\text{DCT}\{\mathbf{p}\}^2[k]\nonumber\\
    &=\frac{\sigma_z^2}{N} \sum_{n=0}^{N-1}\mathbf{p}^2[n]=\sigma_z^2P_p
\end{align}
The second equality comes from the definition in \eqref{eq: p_signal}; the last one is Parseval's theorem. 

\rev{
Finally, the power term $P_p$ can be computed as
\begin{align}
    P_p&=\frac{1}{N}\sum_{n=0}^{N-1}\left(
    \prod_{k=1}^K \left(
    \beta\cos_{m_k}[n]+1
    \right)\right)^2\nonumber\\
    &\approx\sum_{k=0}^{K}
    {K\choose k}\beta^k=(\beta+1)^K,
\end{align}
where the approximation assumes that the number of identical frequencies is negligible. \hfill$\square$}

\rev{
\subsection{Proof of Theorem \ref{thm:error_probability}}
\label{proof_error_prbability}
The error probability can be computed as follows:
\begin{align}
    P_e&=
    \text{Pr}\left(
    \max_{\ell>\Sigma}\mathbf{d}[i]\geq \mathbf{d}[\Sigma]
    \right)\nonumber\\
    &=1-\text{Pr}\left(
    \max_{\ell>\Sigma}\mathbf{d}[\ell]< A_{\Sigma}+\varepsilon[\Sigma]
    \right)
    \nonumber\\
    &\overset{(\text{a})}{=}1-\prod_{\ell>\Sigma}\text{Pr}\Bigl(
    \mathbf{d}[\ell]< A_{\Sigma}+ \varepsilon[\Sigma]
    \Bigr)
    \nonumber\\
    &\overset{(\text{b})}{=}1-\text{Pr}\Bigl(
    \mathbf{d}[\ell]< A_{\Sigma}+ \varepsilon[\Sigma]
    \Bigr)^{N-\Sigma}
    \nonumber\\
    &=1-\left(1-\text{Pr}\Bigl(
    \mathbf{d}[\ell]\geq A_{\Sigma}+ \varepsilon[\Sigma]
    \Bigr)\right)^{N-\Sigma}
    \nonumber\\
    &\overset{(\text{c})}{=}1-\left(1-Q\left(\frac{A_{\Sigma}}{\sigma_{\varepsilon}}\right)
    \right)^{N-\Sigma}
    \nonumber\\
    &\approx (N-\Sigma)Q\left(\sqrt{\frac{A_{\Sigma}^2}{\sigma_{\varepsilon}^2}}\right),
\end{align}
where (a) comes from the independence of the noisy samples, (b) is because these are identically distributed and (c) is the $Q$-function for the tail distribution. \hfill$\square$
}

\rev{
\subsection{Proof of Theorem \ref{thm:mse}}
\label{thm:mse_proof}
The error probability can be developed as
\begin{align}
    \text{MSE}&=
    \sum_{\ell>\Sigma} 
    Pr(\ell|\Sigma)(\ell-\Sigma)^2
    =
    \sum_{\ell=\Sigma}^{N-1} 
    \frac{P_e}{N-\Sigma}(\ell-\Sigma)^2
    \nonumber\\
    &=
    \frac{P_e}{N-\Sigma}
    \sum_{j=1}^{N-\Sigma-1} 
    j^2
    =\frac{P_e}{6}(2N-2\Sigma-1)(N-\Sigma-1),\nonumber
\end{align}
and the final result is achieved substituting $P_e$ in \eqref{eq:error_probability}. \hfill$\square$
}


\bibliographystyle{IEEEbib}
\bibliography{refs}

\end{document}